\documentclass[10pt,conference]{IEEEtran}
\IEEEoverridecommandlockouts
\makeatletter
\let\NAT@parse\undefined
\newcommand{\newlineauthors}{%
  \end{@IEEEauthorhalign}
  \hfill\mbox{}\par
  \mbox{}\hfill\begin{@IEEEauthorhalign}
}
\makeatother

\usepackage{minted}
\usepackage[numbers,sort&compress]{natbib}
\usepackage{amsmath,amssymb,amsfonts}
\usepackage{algorithmic}
\usepackage{graphicx}
\usepackage{textcomp}
\usepackage{url}
\usepackage{xcolor}
\usepackage{listings}
\usepackage{setspace} % Necessário para o \setstretch
\usepackage{tabularx}
\usepackage{booktabs}

\def\BibTeX{{\rm B\kern-.05em{\sc i\kern-.025em b}\kern-.08em
    T\kern-.1667em\lower.7ex\hbox{E}\kern-.125emX}}

\setminted{
    frame=lines,
    framesep=2mm,
    breaklines=true,      
    autogobble=true,     
    fontsize=\footnotesize,
    breaksymbol={},      
    style=bw        
}

\begin{document}

\newcommand{\agentsmd}{\texttt{AGENTS.md}}
\newcommand{\claudemd}{\texttt{CLAUDE.md}}

\definecolor{summaryGray}{RGB}{242, 242, 242}
\definecolor{borderSummaryGray}{RGB}{64, 64, 64}

\newcommand{\mybox}[1]{
  \vspace{0.5em}
  \noindent
  { % Inicia um grupo local para a mudança não vazar
    \setlength{\fboxrule}{0.2pt}
    \fcolorbox{borderSummaryGray}{summaryGray}{\parbox{.975\columnwidth}{#1}}
  } % Termina o grupo local (o \fboxrule volta ao normal)
  \vspace{0.5em}
}

\definecolor{summaryYellow}{RGB}{255, 252, 230}
\definecolor{borderSummaryYellow}{RGB}{160, 160, 160}

\newcommand{\myyellowbox}[1]{
  \vspace{0.5em}
  \noindent
  {
    \setlength{\fboxrule}{0.2pt}
    \fcolorbox{borderSummaryYellow}{summaryYellow}
    {\parbox{.975\columnwidth}{#1}}
  }
  \vspace{0.5em}
}
\newcommand{\red}[1]{\textcolor{red}{#1}}
\newcommand{\joao}[1]{\textcolor{blue}{\textbf{João says:} #1}}
\newcommand{\helio}[1]{\textcolor{orange}{\textbf{Helio  fala:} #1}}

\title{Configuration Smells in AGENTS.md Files: Common Mistakes in Configuring Coding Agents}

\author{\IEEEauthorblockN{Hélio Victor F. dos Santos}
\IEEEauthorblockA{\textit{Department of Computer Science} \\
\textit{Federal University of Minas Gerais}\\
Belo Horizonte, Minas Gerais, Brazil \\
helio.santos@dcc.ufmg.br}
\and
\IEEEauthorblockN{Vitor Costa}
\IEEEauthorblockA{\textit{Department of Computer Science} \\
\textit{Federal University of Minas Gerais}\\
Belo Horizonte, Minas Gerais, Brazil \\
vitorcosta@dcc.ufmg.br}
\and
\IEEEauthorblockN{João Eduardo Montandon}
\IEEEauthorblockA{\textit{Department of Computer Science} \\
\textit{Federal University of Minas Gerais}\\
Belo Horizonte, Minas Gerais, Brazil \\
joao@dcc.ufmg.br}
\newlineauthors
\IEEEauthorblockN{Luciana Lourdes Silva}
\IEEEauthorblockA{\textit{Department of Computing} \\
\textit{Federal Institute of Minas Gerais}\\
Ouro Branco, Minas Gerais, Brazil \\
luciana.lourdes.silva@ifmg.edu.br}
\and
\IEEEauthorblockN{Marco Tulio Valente}
\IEEEauthorblockA{\textit{Department of Computer Science} \\
\textit{Federal University of Minas Gerais}\\
Belo Horizonte, Minas Gerais, Brazil \\
mtov@dcc.ufmg.br}
}

\maketitle

\begin{abstract}
  Coding agents are increasingly used to automate software engineering tasks. To guide their behavior, these agents commonly rely on configuration files, typically named \agentsmd\ or \claudemd, which provide instructions about architecture, workflows, coding conventions, and testing practices. Despite their growing importance, little is known about common problems affecting the definition and maintenance of these files. In this paper, we present the first catalog of smells for coding-agent configuration files. To identify such smells, we first  conducted a grey literature review and a repository mining analysis. As a result, we identified six configuration smells and proposed automated heuristics to detect them.
  To evaluate the prevalence of the proposed smells, we analyzed 100 popular open-source repositories containing either an \agentsmd\ or a \claudemd\ file. Our results show that configuration smells are widespread. {\em Lint Leakage} was the most common smell, affecting 62\% of the files, followed by {\em Context Bloat} (42\%) and {\em Skill Leakage} (35\%). We further show that several smells frequently co-occur, particularly {\em Context Bloat}, {\em Skill Leakage}, and {\em Conflicting Instructions}.

\end{abstract}

% \begin{IEEEkeywords}
% component, formatting, style, styling, insert
% \end{IEEEkeywords}

\section{Introduction}

Large Language Models (LLMs) are transforming the way software is developed by automating a wide range of software engineering tasks. For example, LLMs can assist developers with code generation~\cite{Chen2021a,Shin2023}, bug fixing~\cite{Mastropaolo2023}, test creation~\cite{Siddiq2024, Alshahwan2024}, code review~\cite{llama-reviewer}, documentation writing~\cite{ian, SystematicLiteratureReview}, software migration~\cite{aylton, google-migrate-code}, and code smell detection~\cite{ESEM24:silva}. Initially, these capabilities were delivered through coding assistants that interactively support developers while they write code. For example, this was the case with the first version of GitHub Copilot, launched in 2021. More recently, however, automation has advanced with the emergence of coding agents, such as Claude Code\footnote{\url{https://claude.com/product/claude-code}}, Codex\footnote{\url{https://openai.com/codex/}}, Cursor Agent\footnote{\url{https://cursor.com/agents}}, and Gemini CLI\footnote{\url{https://ai.google.dev/gemini-api/docs}}, which can execute complex tasks with limited intervention.

Coding agents are designed to operate autonomously. Given a high-level goal, they can plan and execute multiple actions until the requested task is completed. Architecturally, a coding agent can be viewed as a combination of a language model and a harness. The language model provides reasoning and inference capabilities, while the harness implements an agentic loop that repeatedly interacts with the model, executes actions, and feeds the results back to the model. During this loop, agents can invoke a variety of external tools, including code search utilities, shell commands, test runners, version-control systems, web search engines, and issue-tracking platforms.

To guide their behavior, coding agents commonly rely on project-specific configuration files, typically named \agentsmd\ or \claudemd. These files contain instructions that complement the agent's built-in capabilities, such as coding conventions, architectural guidelines, testing requirements, project workflows, and domain-specific knowledge. Their main purpose is to provide persistent contextual information that helps agents behave consistently across different tasks and sessions. In most agent harnesses, the configuration files are loaded when a session starts, incorporated into the agent's prompt, and maintained as part of the context available throughout the execution of the agentic loop.

Given their importance for the performance of coding agents, {\bf in this paper we present a catalog of smells commonly found in agent configuration files}. To identify these smells, we conducted a review of the grey literature, covering 14 recent articles on the topic. As a result, we identified six smells that may occur in these files and, consequently, impair the overall performance of coding agents. We then used a set of heuristics to detect these smells in a dataset of 100 popular open-source projects containing either an \agentsmd\ or a \claudemd\ file. In total, we identified 207 instances of the proposed smells. The most common smell was {\em Lint Leakage}, with 62 identified instances. This smell refers to instructions in configuration files that essentially restate rules already enforced by automated tools such as linters and formatters, thus unnecessarily consuming context space and tokens.
We also analyzed the co-occurrence of smells in our dataset and discovered several strong relationships. For example, two smells ({\em Skill Leakage} and {\em  Conflicting Instructions}) increase the likelihood of {\em Context Bloat} by 83\%. 
% These co-occurrence relationships are important because they suggest that certain smells should be prioritized for removal, as eliminating them may consequently eliminate other smells as well.
These results are important because they allow us to understand which smells can trigger the appearance of others in configuration files.

The remainder of this paper is organized as follows. Section~\ref{sec:coding-agents} provides background on the role and importance of configuration files in agentic development. Section~\ref{sec:methods} details our study design and methods, which included a grey literature review, the creation of a dataset of real-world agent configuration files, and a manual analysis of pull requests involving changes to such files. Section~\ref{sec:configuration-smells} describes the configuration smells that can occur in agent configuration files, i.e., the proposed catalog. Next, Section~\ref{sec:heuristics} defines the heuristics used to detect these smells, and Section~\ref{sec:smells-wild} reports their occurrence in our dataset. Section~\ref{sec:cooccurrence} then investigates co-occurrence relationships among the identified smells. Finally, Section~\ref{sec:threats} discusses threats to validity, Section~\ref{sec:related-work} reviews related work, and Section~\ref{sec:conclusion} concludes the paper.

% \clearpage

\section{Coding Agents Configuration Files}
\label{sec:coding-agents}

While the effectiveness of coding agents is heavily dependent on the capacity of the underlying model, it is equally restricted by the harness used for task execution. In this scenario, agent configuration files—such as \agentsmd\ and \claudemd—emerge as a standard for context injection and for providing a persistent memory for coding agents. According to~\citet{xi2023risepotentiallargelanguage}, the effectiveness of an agent lies not only in the power of the base model but in the precision with which guidelines are defined, stored, and retrieved.

In essence, \agentsmd\ is a markdown file with rules about the project.
In this file, one can add information regarding system architecture, tool documentation, testing practices, and other constraints whose absence might lead to agent errors.
In a recent study,~\citet{agent2026-claude-code} indicate that the most common sections in these files cover architecture, development rules, project overviews, and testing workflows.

Upon starting a session, the agent detects the presence of \agentsmd\ files in the project directory and incorporates their content into the agent's prompt, persisting this context throughout the entire agentic loop~\cite{mohsenimofidi2025context}.
An example of a configuration file can be seen in Figure \ref{fig:agent-md}.\footnote{This example was extracted and adapted from \url{https://agents.md}} This file includes a commands section for project setup, project-specific code style rules, and workflow definitions for the model.

\begin{figure}[ht]
  \begin{minted}{markdown}
# AGENTS.md
 
## Setup commands
- Install deps: `pnpm install`
- Start dev server: `pnpm dev`
- Run tests: `pnpm test`
 
## Code style
- TypeScript strict mode
- Use functional patterns where possible
- Use ES modules (import/export) syntax, not CommonJS (require)

## Workflow
- Be sure to typecheck when you're done making a series of code changes
- Prefer running single tests, and not the whole test suite, for performance
    
  \end{minted}
  \caption{Example of \agentsmd\ file}
  \label{fig:agent-md}
\end{figure}

The first version of the \agentsmd\ file can be created automatically by the agent itself.
In this case, a specific command---generally \texttt{/init}---relies on an internal prompt instructing the agent to inspect the repository and leverage an initial set of guidelines.
For example, the prompt used by the OpenAI Codex---which is publicly available\footnote{\url{https://github.com/openai/codex/blob/main/codex-rs/tui/prompt_for_init_command.md}}---asks the agent to summarize the project structure and module organization, build and testing commands, coding and naming conventions, testing guidelines, and pull request requirements, while keeping the resulting document concise, instructional, and tailored to the analyzed repository.

It is also possible to use both files in the same repository. In this case, one file should simply point to the other one. For example, the \claudemd\ file may exist but contain only the following line: {\tt read \@AGENTS.MD}.

\myyellowbox{{\em Note:} Most agent-based systems use the name \agentsmd\ for their configuration files. The notable exception is Claude Code, which uses the name \claudemd. Therefore, in this paper, we analyze both \agentsmd\ and \claudemd\ files, since the only difference between them, in terms of purpose and role within an agent-based system, is their name.
}

\section{Methods}
\label{sec:methods}

To research the smells that affect agent configuration files, we employed three methods: a grey literature review (Section~\ref{sec:grey-literature}), the creation of a dataset of real-world agent configuration files (Section~\ref{sec:dataset}), and an analysis of discussions conducted in pull requests (Section~\ref{sec:pull-requests}).

\subsection{Literature Review}
\label{sec:grey-literature}

\begin{figure}[h]
  \centering
  \includegraphics[width=\columnwidth]{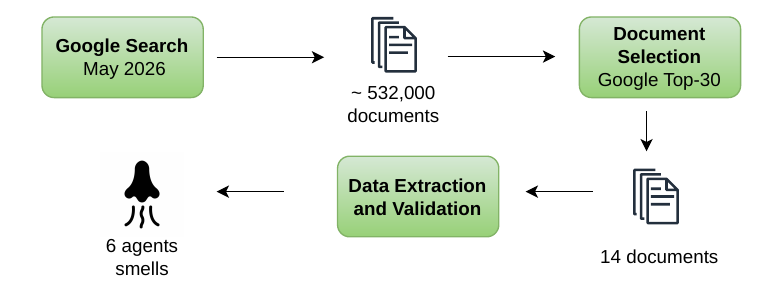}
  \caption{Overview of the grey literature steps}
  \label{fig:h2_titles}
\end{figure}

Grey literature review is commonly used in software engineering research to gather information for emerging topics, since it emphasizes the inclusion of non-academic sources, such as blog posts, technical reports, and documentation~\cite{garousi2018guidelinesincludinggreyliterature}.
Our procedure is depicted in Figure~\ref{fig:h2_titles} and consists of three main steps: (a) Google search, (b) Document selection, and (c) Data extraction and validation.\\[-0.3cm]

\noindent{\em Document Search:}
We started by performing a Google search to identify documents that describe bad practices or smells in \agentsmd\ files.
Similar to other grey literature reviews, we restricted our search to Google due to its wide coverage and relevance for retrieving non-academic sources~\cite{lucasSmells, google-search}.
Specifically, the search query, as presented in Figure \ref{fig:search-string}, looked for documents containing the terms {\em agents.md} or {\em claude.md} together with terms related to {\em bad smells}, {\em anti-patterns}, and {\em best practices}.\\[-0.3cm]

\begin{figure}[ht]
  \begin{minted}{sql}
  ("agents.md" OR "AGENTS.md" OR "claude.md" OR "CLAUDE.md") 
  AND 
  ("bad smell" OR "bad smells" OR "anti-pattern" OR "anti-patterns" OR "antipattern" OR "antipatterns" OR "bad-practice" OR "bad-practices" OR "bad practice" OR "bad practices" OR "best-practices" OR "best practice" OR "best practices")
  \end{minted}
  \caption{Search string used in the literature review}
  \label{fig:search-string}
\end{figure}

\noindent{\em Document Selection:}
Our search returned 532,000 documents.
From these results, we manually analyzed the first 30 documents returned by the search; i.e.,~the ones returned in the first three pages.
The first author discarded 16 documents according to the following criteria: 
(a) five documents were accessible only through subscription or paywall; 
(b) four documents were out of scope, i.e.,~focused on specific tools or other topics; 
(c) three represented discussion threads on forums; 
(d) three documents did not explain any bad smell or best practice on \agentsmd\ files, they only listed these issues pointing to the official documentation of the tools; and 
(e) one document was written in a non-English language.
As a sanity check, the third author analyzed each discarded document to confirm the reason for exclusion; no disagreements were found in this process. 
The remaining {\bf 14 documents were selected for the next step}. We also assessed the quality and reputation of the authors of all selected documents. Six documents were published by well-known companies, such as Anthropic and GitHub. The remaining eight documents were written by authors with relevant expertise in the field (seven authors have at least six years of experience, and one author has two years). Furthermore, three of the articles were extensively discussed on Hacker News (a well-known forum for technology and startup discussions), receiving at least 140 upvotes each. Throughout this paper, we refer to these documents using the identifier $D_n$, where $n$ is a numeric identifier. The list of selected documents is available in our replication package.\\[-0.3cm]

\noindent{\em Smells Extraction and Validation:}
For each selected document, the first author thoroughly read its content and marked any sentence that described practice to avoid when creating \agentsmd\ files.
Following this initial extraction, the second author reviewed the annotations as a sanity check to confirm whether they actually describe a bad smell. This process revealed four disagreements, which both authors discussed to reach a consensus. After this annotation process, the first and last authors manually grouped sentences describing similar smells and proposed a name and a short description for each group. They also decided to discard one smell that is not specific to coding agents: exposing secret keys in configuration files. At the end, {\bf we obtained a list of six smells}.

\subsection{Dataset of Agent Configuration Files}
\label{sec:dataset}

To investigate whether the  smells actually occur in real projects, we created {\bf a dataset with 100 agent configuration files}, including 39 \agentsmd\ and 61 \claudemd\  files. We selected these files using the following steps.
In January 2026, we used the GitHub Search API to find repositories containing either \agentsmd\ or \claudemd\ in their root folder. 
Next, we manually removed from this list repositories that are not applications, e.g.,~awesome lists and tutorials. From the remaining ones, we selected the top-100 repositories with the highest number of stars.
For example, our dataset of selected GitHub projects includes \texttt{n8n-io/n8n} (a workflow automation platform)\footnote{\url{https://github.com/n8n-io/n8n}}, \texttt{langchain-ai/langchain} (an agent engineering platform)\footnote{\url{https://github.com/langchain-ai/langchain}}, and \texttt{vercel/ai} (an AI Toolkit for TypeScript).\footnote{\url{https://github.com/vercel/ai}}

\subsection{Pull Requests Analysis}
\label{sec:pull-requests}

After creating the dataset of files described in the previous section, we decided to inspect the Pull Requests (PRs) of the corresponding projects to determine whether they contained discussions about potential issues in agent configuration files. Such issues could help expand the list of six smells identified through our grey literature review. Next, we explain how the pull requests were selected and analyzed.\\[-0.3cm]

\noindent{\em Pull Requests Selection:} 
We collected all commits associated with each of the 100 configuration files in our dataset, including modifications, additions, and deletions. 
This process yielded an initial dataset composed of 760 commits.
From this initial dataset, we selected only the ones linked to pull requests, resulting in a total of 383 pull requests.

A manual inspection was then performed on the conversation history of each filtered pull request to check whether the ongoing discussions about the target files, \claudemd\ and \agentsmd, were in fact of substantive relevance. We discarded all PRs that met any of the following exclusion criteria: 
(a) conversations consisting entirely of bot-generated messages; 
(b) threads where human participants interacted exclusively with bots; 
(c) discussions that failed to actively engage with or reference the \agentsmd\ or \claudemd\ files; 
(d) PRs containing only the initial opening description without any subsequent discussion or peer review.
The application of these sequential filters yielded a {\bf final corpus of 17 relevant pull requests}. \\[-0.3cm]

\noindent{\em Pull Requests Analysis:} After analyzing the 17 pull requests, we found discussions about configuration-related problems in only two of them. These pull requests were labeled $PR_1$ and $PR_2$ and are also included in our replication package. However, the discussions concerned smells that had already been identified in the grey literature review. Therefore, {\bf our list remained unchanged, containing six smells}. Although no new smells were discovered, the analysis of PRs was still valuable because it provided concrete examples of discussions about smells in agent configuration files, which we use to illustrate the content of Section~\ref{sec:configuration-smells}.\footnote{As an additional note to this analysis, in six PRs, the goal was to consolidate agent-related documentation into a single file. For example, one PR was described as a pure rename of \claudemd\ to \agentsmd.}

\section{Configuration Smells}
\label{sec:configuration-smells}

Table~\ref{tab:smells-config} presents a list of the smells identified in our study.
The table also shows the identifiers of the articles in which these
smells were found. In the remainder of this section, we provide a description of the smells listed in this table.

\begin{table}[ht]
  \centering
  \caption{Configuration Smells}
  \label{tab:smells-config}
  \begin{tabularx}{\columnwidth}{l >{\raggedright\arraybackslash}X}
    \toprule
    \textbf{Smells}          & \multicolumn{1}{c}{\textbf{Articles and PRs}}      \\ \midrule
    Context Bloat            & $D_{1}$, $D_{3}$, $D_{4}$, $D_{5}$, $D_{6}$, $D_{9}$, $D_{11}$, $D_{12}$, $D_{13}$, $D_{14}$, $PR_1$ \\
    Skill Leakage            & $D_{2}$, $D_{3}$, $D_{4}$, $D_{5}$, $D_{6}$, $D_{11}$, $D_{12}$, $D_{13}$, $PR_2$          \\
    Lint Leakage             & $D_{1}$, $D_{4}$, $D_{5}$, $D_{6}$, $D_{9}$, $D_{13}$                    \\
    Blind Reference          & $D_{6}$, $D_{7}$                                     \\
    Init Fossilization       & $D_{4}$, $D_{6}$, $D_{9}$, $D_{10}$, $D_{13}$                       \\
    Conflicting Instructions & $D_{1}$, $D_{3}$, $D_{6}$                                 \\ \bottomrule
  \end{tabularx}
\end{table}

\subsection{Context Bloat}\label{config-context-bloat}

This was the most frequently cited smell in our documents, being mentioned in 10 out of the 14 reviewed articles and in one PR. It occurs when an \agentsmd\ file becomes excessively large and overloaded with rules, examples, or low-priority details. Bloated configuration files increase token consumption, raise costs, and reduce the visibility of important instructions. Therefore, \agentsmd\ files should remain concise and focused on essential project-specific guidance. For example, Anthropic's documentation explicitly recommends the following: {\em target under 200 lines per \claudemd\ file. Longer files consume more context and reduce adherence.}\footnote{\url{https://code.claude.com/docs/en/memory}}

To provide another example, in one of the pull requests we analyzed ($PR_1$), the authors proposed restructuring the \agentsmd\ file to reduce its size. The PR explicitly states that the project's configuration file was reduced from 598 to 149 lines, because {\em modern LLMs tend to perform better with configuration files containing approximately 150 to 200 lines}.

\subsection{Skill Leakage}

This smell occurs when specific, rarely used, or highly
context-dependent instructions are placed in the \agentsmd\ file instead of being specified in dedicated skill files (e.g.,~{\tt skills.md}) and loaded on demand. In practice, this means that specialized knowledge “leaks” into every agent session, even when it is not needed. As a result, the agent's context becomes larger, more expensive, and harder to maintain. Furthermore, such rules may compete for attention with the rules that are actually critical for the project. For example, one of the articles explicitly recommends the following: {\em Instead of including all your different instructions about building your
project, running tests, code conventions, or other important context in your
CLAUDE.md file, we recommend keeping task-specific instructions in separate
markdown files with self-descriptive names somewhere in your project.} ($D_{13}$)

%{\em Treat your \agentsmd\ like a skill. Cover the common cases and workflows at a high level, then push details into reference files the agent can load on demand. Keep each reference's scope clear so the agent knows when to pull it in.}. ($D_{11}$)

%\helio{tem uma frase que é razoavelmente grande mas ela cobre bem, somente se adaptarmos ela "Writing a concise CLAUDE.md file that covers everything you want Claude toknow can be challenging, especially in larger projects.[paragrafo] To address this, we can leverage the principle of Progressive Disclosure to ensure that claude only sees task- or project-specific instructions when it needs them. [paragrafo] }

\subsection{Lint Leakage}

This smell occurs when an \agentsmd\ file includes rules that are already checked by linters, formatters, or other static analysis tools. Typical examples include naming conventions (such as camelCase or PascalCase), formatting rules, import ordering, maximum line length, or generic style-guide recommendations. Because these constraints are automatically checked by local tools, repeating them in \agentsmd\ adds limited value while unnecessarily increasing the agent's context size~\cite{Huyen2025}. Moreover, emphasizing such coding rules can divert the model from focusing on more important project-specific concerns, such as architectural constraints, domain rules, or safety policies. For example, one of the articles explicitly recommends the following: {\em Code style enforcement is the biggest trap. Formatting, indentation, import ordering: these are deterministic problems with deterministic solutions. Linters and formatters like Biome, ESLint, or Ruff handle them faster, cheaper, and with 100\% consistency. Spending instruction budget on style rules is dead weight: the same work a pre-commit hook does for free}. ($D_{6}$)

\subsection{Blind References}

This smell occurs when an \agentsmd\ file contains references to external documents, files, or directories without explaining their purpose or scope. As a consequence, the agent may unnecessarily load large documents into context, ignore important references, or fail to prioritize the correct source of information for a given task. Thus, a better practice is to complement references with concise descriptions explaining the role of the document, the type of information it contains, and the context in which it should be used. For example, one document explicitly recommends the following: {\em If you just mention the path [of an external document], Claude will often ignore it. You have to pitch the agent on why and when to read the file}. ($D_{7}$)

\subsection{Init Fossilization}

This smell occurs when the configuration file is generated by an initialization command such as {\tt /init} but not reviewed or updated afterwards.
Thus, the file generated by the coding agent becomes the permanent configuration, often carrying instructions that are not relevant anymore to the project. As a result, the configuration tends to accumulate noise, increase context consumption, and reduce the overall effectiveness of the agent over time. For example,
Anthropic’s documentation explicitly recommends that the configuration file should be continuously updated, such as when: {\em Claude makes the same mistake a second time;
a code review reveals something Claude should already have known about the codebase;
you find yourself typing the same correction or clarification in chat that you already provided in a previous session; and a new team member would need the same context in order to be productive.}\footnote{\url{https://code.claude.com/docs/en/memory}}

%$D_{13}$ document explicitly recommends the following: {\em Don’t use /init or let the agent write its own AGENTS.md. The data shows this hurts more than it helps}.

\subsection{Conflicting Instructions}

This smell occurs when an \agentsmd\ file contains instructions
that contradict each other, creating ambiguity about the expected behavior of the agent. Such inconsistencies can confuse the model and lead to unstable results. For example, small variations in the prompt can result in different agent behavior, because they are enough to make the agent follow a different inference path than the one followed previously. Anthropic’s documentation explicitly recommends the following: {\em if two rules contradict each other, Claude may pick one arbitrarily. Review your CLAUDE.md files periodically to remove outdated or conflicting instructions.}\footnote{\url{https://code.claude.com/docs/en/memory}}

%$D_{6}$ document explicitly recommends the following: {\em "Review this CLAUDE.md and suggest improvements" every few weeks. It spots contradictions between rules, flags overlapping instructions, and identifies phrasing that could be tighter}.

%\helio{...If Claude asks you questions that are answered in CLAUDE.md, the phrasing might be ambiguous. Treat CLAUDE.md like code: review it when things go wrong, prune it regularly, and test changes by observing whether Claude’s behavior actually shifts. ($D_{1}$)}

\section{Detection Heuristics}
\label{sec:heuristics}

We also propose a set of heuristics to detect the smells described in the previous section.

\subsection{Heuristic based on Lines of Code}

This heuristic is used exclusively to detect the {\em Context Bloat} smell. Specifically, we decided to use a threshold of 200 lines of code to identify this smell, as also suggested in the Anthropic document mentioned in Section \ref{config-context-bloat}. In other words, \agentsmd\ files with 200 or more lines of code are classified as presenting the {\em Context Bloat} smell.

\subsection{Heuristic based on Language Models }

To detect the {\em Skill Leakage}, {\em Lint Leakage}, {\em Blind References}, and {\em Conflicting Instructions} smells, we decided to use a large language model. We designed a dedicated prompt for each smell, thereby requesting a more specific and objective task from the model. Figure~\ref{fig:example-prompt} illustrates the baseline template of these prompts. 

\begin{figure}[ht]
  \begin{minted}{markdown}
# Context

You are a senior software engineer. I will provide
an agent configuration file (AGENTS.md). Your task 
is to detect whether the file contains the 
following configuration smell.

# Smell: [name]

Description: [as in Section IV; essentially the first
paragraph of the smell description]

[Optional: Operational Guidelines / True & False Positives Examples]
- Examples and edge cases specific to the smell to minimize false positives.

# Output

- If detected: [Return exact lines / JSON object with contradiction context]
- If not detected, return only: NO SMELL

# Agents.md file

[file content]
  \end{minted}
  \caption{Prompt to detect Skill Leakage, Lint Leakage, Blind References and Conflicting Instructions}
  \label{fig:example-prompt}
\end{figure}

In the case of {\em Blind References} and {\em Conflicting Instructions}, we also enriched the basic prompt with examples of true/false positives. For example, in  Figure \ref{fig:blind-references-example-heuristics} we show the example (true positive) added to the 
prompt to detect {\em Blind References}.

\begin{figure}[ht]
  \begin{minted}{markdown}
...
## Releases & Environment

For releases or environment issues, see
`web/book/src/project/contributing/development.md`.
\end{minted}
  \caption{Example added to the prompt to detect Blind References}
  \label{fig:blind-references-example-heuristics}
\end{figure}

\subsection{Heuristic based on Number of Commits}

This heuristic is used exclusively to detect the {\em Init Fossilization} smell. Conservatively, we consider that an \agentsmd\ file with only a single commit exhibits this smell; that is, the file has never been modified since its creation.

\section{Configuration Smells in the Wild}
\label{sec:smells-wild}

To detect the smells in our dataset of 100 agent configuration files, we applied the heuristics proposed in Section~\ref{sec:heuristics}. For the heuristics based on LLMs, we used gemini-3.1-flash-lite, with a temperature of 0. Additionally, each smell instance identified by these heuristics was carefully reviewed by the first author to verify whether it is a true occurrence of the smell. The results are summarized in Table~\ref{tab:smells-result} and are discussed in the following subsections.

\begin{table}[ht]
  \centering
  \caption{Smells detected in real projects \\(FP: False Positives; Prec: Precision)}
  \label{tab:smells-result}
  \begin{tabularx}{\columnwidth}{X c c c}
    \toprule
                             & \textbf{Instances} & \textbf{FP} & \textbf{Prec. (\%)} \\ \midrule
    Context Bloat            & 42                 & -           & -                   \\
    Skill Leakage            & 35                 & 6           & 82                  \\
    Lint Leakage             & 62                 & 4           & 93                  \\
    Blind Reference          & 16                 & 2           & 87                  \\
    Init Fossilization       & 24                 & -           & -                   \\
    Conflicting Instructions & 28                 & 12          & 57                  \\ \bottomrule
  \end{tabularx}
\end{table}

In Table~\ref{tab:smells-result}, we also show the number of false positives and the precision, for the smells whose identification relies on LLM-based heuristics. In other words, we did not compute precision for {\em Context Bloat} and {\em Init Fossilization} because, in these cases, detection is based on pre-established thresholds.

\myyellowbox{{\em Summary:} We detected at least one smell in 91 agent configuration files. Thus, only nine files were found to be smell-free. These results suggest that developers could benefit from catalogs and tools designed to spot  configuration issues in agent configuration files.}

\subsection{Context Bloat}

The proposed heuristic detected 42 cases of {\em Context Bloat}. The smallest file contains 216 lines of code, whereas the largest file contains 1,477 lines of code.  Due to space constraints, we will not present a complete example of {\em Context Bloat} here. However, just to provide a high-level illustration, the \claudemd\ file of the {\tt javascript-obfuscator} project---a very popular and powerful obfuscator of JavaScript and Node.js source code---has 1,477 lines.\footnote{\url{https://github.com/javascript-obfuscator/javascript-obfuscator}} This file is organized into 27 sections, including Project Overview, Architecture Overview, Core Workflow, CLI/API Usage, etc. As a result, the file is very large for repeated context loading, which can lead language models to discard important instructions. Much of these instructions would be better maintained in separate documentation or loaded on demand through skills.

When analyzing the file, we noticed, for example, that the second section is called {\em Key Features}, and has 22 lines. This section describes the obfuscation techniques used by the project, such as variable and function renaming, string extraction and encryption, dead code injection, and control-flow flattening. This information is mostly product documentation and provides limited value as persistent context for agents. Therefore, it could be removed from the agent configuration and described in the project's README, for example.

It is also important to note that {\em Context Bloat} is a more visible smell, which makes it easier to detect. The root cause of this smell is the presence of other smells in the configuration file, such as {\em Skill Leakage}, which we will discuss next.

\subsection{Skill Leakage}

The proposed heuristic detected 35 cases of {\em Skill Leakage}. After a manual analysis conducted by the first author, 29 cases were confirmed (86\%).  An example of {\em Skill Leakage} was found in the \agentsmd\ file of {\tt quickemu-project/quickemu}, which is a tool to simplify the creation and execution of virtual machines.\footnote{\url{https://github.com/quickemu-project/quickemu}} The detected smell instance is shown in Figure~\ref{fig:skill-leakage}. As we can see, the section {\em Adding a new OS to quickget} contains instructions that are only useful for a small subset of tasks. Since most interactions with the coding agent do not involve adding new operating systems, these instructions unnecessarily increase the size of the configuration file. Thus, they would be better placed in a dedicated skill or documentation file.\\[-0.3cm]

\begin{figure}[ht]
  \begin{minted}{markdown}
## Adding a new OS to quickget

Follow the [guide in the wiki](...). Each OS requires:

1. Entry in `os_info()` case statement
2. `releases_<os>()` function returning available versions
3. `editions_<os>()` function if multiple editions exist
4. `arch_<os>()` function if ARM64 is supported (defaults to amd64 only if omitted)
5. Download URL construction logic
  \end{minted}
  \caption{Example of Skill Leakage (quickemu-project/quickemu)}
  \label{fig:skill-leakage}
\end{figure}

\noindent{\em Most common leaked skills:} We also manually classified the skills responsible for the identified {\em Skill Leakage} instances. The results of this classification are presented in Table~\ref{tab:skill-type}. As can be seen, the most common skills incorrectly defined in \agentsmd\ files are related to testing concerns, followed by workflow guidelines (e.g., procedures for code reviews, pull requests, and issue management). Particularly, the Skill Leakage instance presented in Figure~\ref{fig:skill-leakage} was classified as scaffolding, as it defines functions that must be implemented to support a new operating system image within a module.

\begin{table}[ht]
  \centering
  \caption{Most common leaked skills}
  \label{tab:skill-type}
  \begin{tabular}{l c}
    \toprule
    \textbf{Skill Type} & \textbf{Frequency} \\ \midrule
    Testing             & 10                 \\
    Workflow            & 8                  \\
    Scaffolding         & 4                  \\
    Infrastructure      & 4                  \\
    Architecture        & 3                  \\
    \bottomrule
  \end{tabular}
\end{table}

\subsection{Lint Leakage}

The proposed heuristic detected 62 cases of {\em Lint Leakage}. After a manual analysis conducted by the first author, 58 cases were confirmed (93\%). An interesting case of this smell was identified in the {\tt google/adk-python} project, which is an open-source SDK for building agent-based applications in Python.\footnote{\url{https://github.com/google/adk-python}}
As shown in Figure~\ref{fig:lint-leakage}, the \agentsmd\ file of this project includes a section called {\em Python Style Guide} containing instructions for writing Python code, including recommendations for indentation and line length, naming conventions, usage of docstrings, etc. Normally, these recommendations are enforced by linters, formatters, or widely adopted community conventions, making their inclusion in the file unnecessary.

\begin{figure}[ht]
  \begin{minted}{markdown}
### Python Style Guide

* Indentation: 2 spaces.
* Line Length: Maximum 80 characters.
* Naming Conventions**:
  * `function_and_variable_names`: `snake_case`
  * `ClassNames`: `CamelCase`
  * `CONSTANTS`: `UPPERCASE_SNAKE_CASE`
* Docstrings: Required for all public modules, ...
* Imports: Organized and sorted.
* Error Handling: Specific exceptions should be ...
\end{minted}
  \caption{Example of Lint Leakage (google/adk-python)}
  \label{fig:lint-leakage}
\end{figure}

However, after creating our dataset (in January, 2026), we found that the project maintainers performed a major refactoring of the \agentsmd\ file. Specifically, the {\em Python Style Guide} section,  shown in Figure~\ref{fig:lint-leakage}, was moved to a separate skill file.
Therefore, this extraction confirms the relevance of the smell we initially detected in our dataset.\\[-0.3cm]

\subsection{Blind Reference}

To help explain this smell, Figure~\ref{fig:blind-references-no-smell} shows an example in which an external reference is cited appropriately. Notice that the text references an external dependency, includes a link to its GitHub repository, and provides a brief explanation of its purpose ({\em cdp-use only provides shallow typed interfaces for the websocket calls}). Consequently, the agent is able to understand the role of the dependency without needing to load or inspect the external repository directly.

\begin{figure}[ht]
  \begin{minted}{markdown}
## CDP-Use

We use a thin wrapper around CDP called cdp-use: https://github.com/browser-use/cdp-use. cdp-use only provides shallow typed interfaces for the websocket calls, all CDP client and session management + other CDP helpers still live in browser_use/browser/session.py.
\end{minted}
  \caption{External reference described with context (browser-use/browser-use)}
  \label{fig:blind-references-no-smell}
\end{figure}

However, after applying the proposed heuristic, we were able to detect 16 instances of  {\em Blind Reference}, i.e., references cited in \agentsmd\ files without appropriate context. After a manual analysis, 14 cases were confirmed (87\%). An example is shown in Figure~\ref{fig:blind-references}. As we can see, this configuration references an external document ({\tt docs/plugin-reorg.md}) to explain the planned plugin system, but it does not provide any contextual information about the document itself. Therefore, the agent would need to load and inspect the referenced file to understand the architecture and goals of the plugin system.

\begin{figure}[ht]
  \begin{minted}{markdown}
### Plugin System (v5.0 - Not Yet Available)

The TypeScript plugin system (`.claude-plugin/`, marketplace) is planned for v5.0.
See `docs/plugin-reorg.md` for details.
...
  \end{minted}
  \caption{Blind Reference (SuperClaude-Org/SuperClaude\_Framework)}
  \label{fig:blind-references}
\end{figure}

\subsection{Init Fossilization}

\begin{figure}[h!]
  \centering
  \includegraphics[width=0.9\columnwidth]{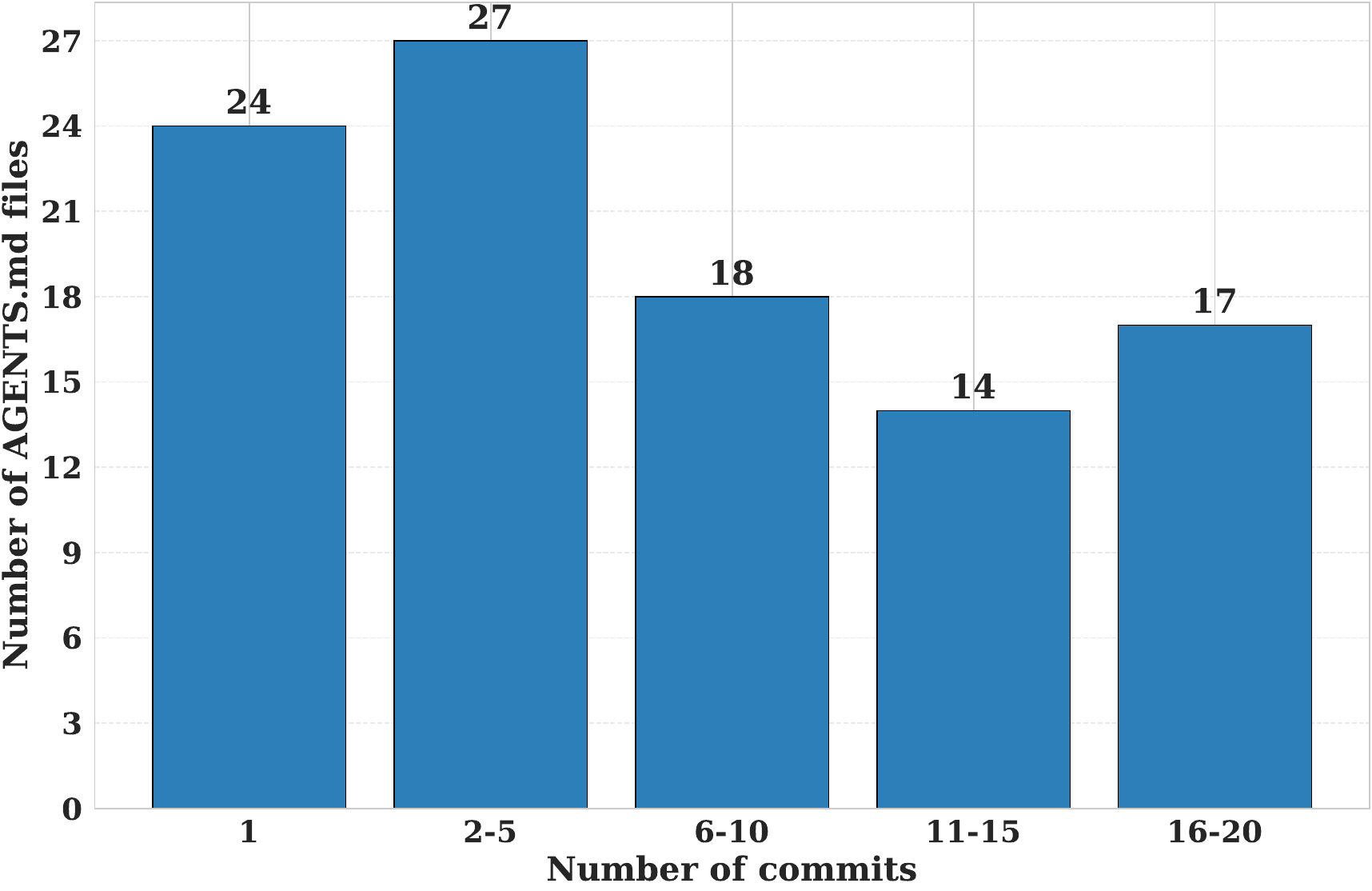}
  \caption{Number of changes in \agentsmd\ files ({\em Init Fossilization} corresponds to files with a single commit)}
  \label{fig:init_fossilization}
\end{figure}

We detected 24 cases of {\em Init Fossilization}, that is, \agentsmd\ files with a single commit, as illustrated by the histogram in Figure~\ref{fig:init_fossilization}. The histogram also shows that configuration files are frequently updated in practice, thus reinforcing our argument that the absence of changes in such files is indeed a smell. For example, 14 of the analyzed files (14\%) have between 11 and 15 commits, while 17 files (17\%) have between 16 and 20 commits.

\begin{figure}
  \centering
  \includegraphics[width=0.9\columnwidth]{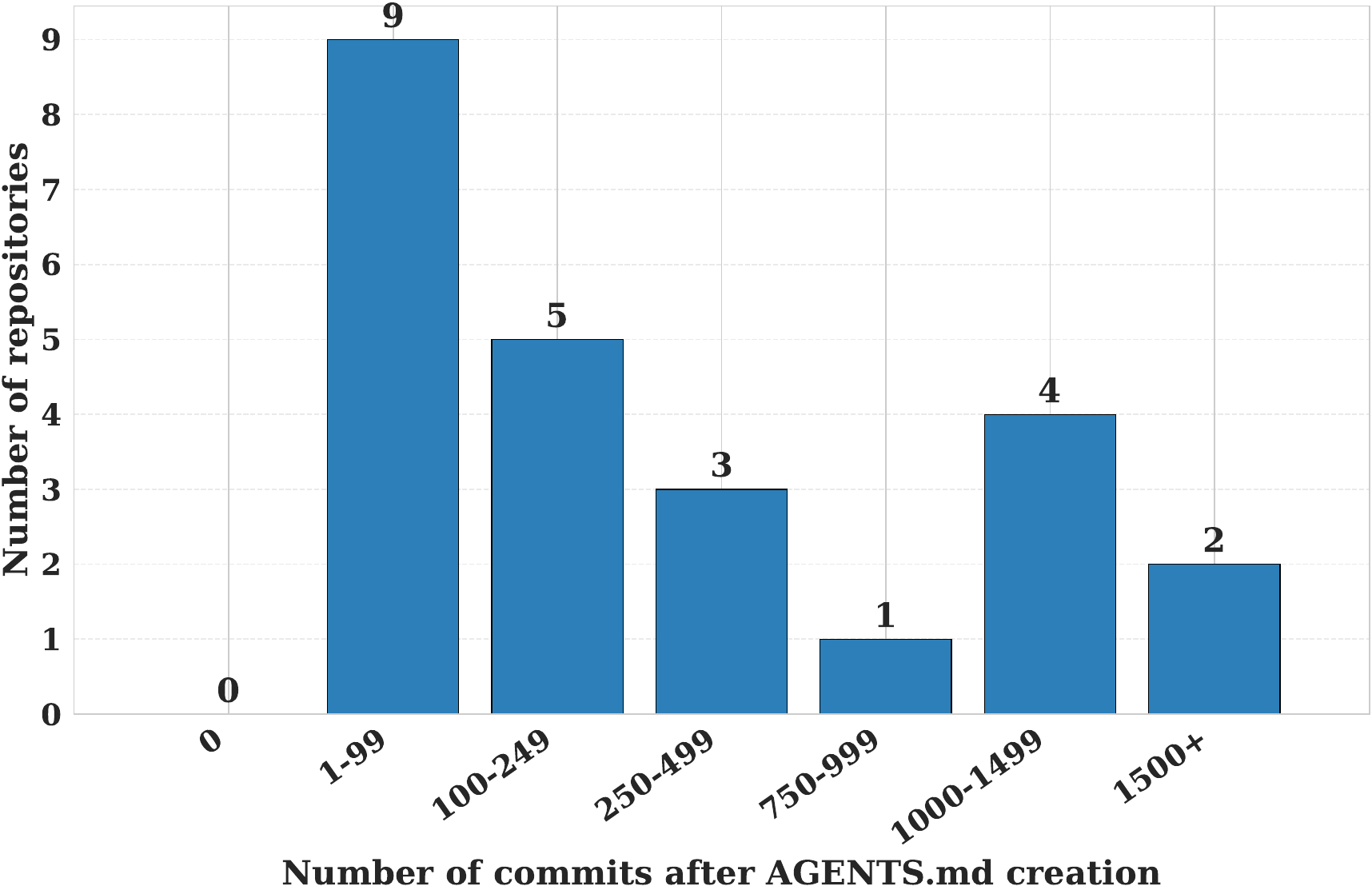}
  \caption{Number of commits in projects exhibiting {\em Init Fossilization} (counting only commits made after the creation of \agentsmd).
  }
  \label{fig:commits_init-fossilization-projects}
\end{figure}

However, it is possible that the 24 projects exhibiting {\em Init Fossilization} were dormant projects, that is, projects with limited activity and few commits. Therefore, the histogram in Figure~\ref{fig:commits_init-fossilization-projects} shows the total number of commits made to these projects after the creation of their respective \agentsmd\ files. As the histogram indicates, the hypothesis that these projects were inactive was not supported. In fact, we did not find a single project in such a situation, that is, with zero commits after the creation of the \agentsmd\ file. On the contrary, we observed many projects with a substantial number of commits, including two projects with more than 1,500 commits, for example.

\subsection{Conflicting Instructions}

The proposed heuristic detected 28 cases of {\em Conflicting Instructions}. However, after a manual analysis conducted by the first author, only 16 cases were confirmed (57\%).
This lower precision is, to some extent, understandable, since identifying contradictory instructions is indeed a more complex task.
Figure~\ref{fig:conflicting-instructions} shows an example of this smell. As can be observed, the configuration specifies two directory paths for creating new components. An agent cannot satisfy the requirement to place components in {\tt packages/ui/components} while simultaneously following the instruction to create them in {\tt packages/components}.

\begin{figure}[ht]
  \begin{minted}{markdown}
    ...
# Component Guidelines
- Components should be placed in the `packages/ui/components` directory
- ...

## How to create a new component

- Create a new folder in `packages/components` with the name of the component.
...
  \end{minted}
  \caption{Example of Conflicting Instructions (inkline/inkline)}
  \label{fig:conflicting-instructions}
\end{figure}

\begin{table*}[t]
  \centering
  \caption{Co-existing smells detected by Apriori.}
  \label{tab:smells-apriori}
  \begin{tabular}{l l l | c c c}
    \toprule
    \multicolumn{2}{l}{\textbf{Antecedent}}                   & \textbf{Consequent} & \textbf{Support}       & \textbf{Confidence} & \textbf{Lift}        \\ \midrule
    \textit{Conflicting Instructions}, \textit{Skill Leakage} & $\rightarrow$       & \textit{Context Bloat} & 0.06                & 0.83          & 1.81 \\
    \textit{Init Fossilization}, \textit{Skill Leakage}       & $\rightarrow$       & \textit{Lint Leakage}  & 0.06                & 0.83          & 1.31 \\
    \textit{Conflicting Instructions}                         & $\rightarrow$       & \textit{Context Bloat} & 0.14                & 0.81          & 1.76 \\
    \textit{Skill Leakage}                                    & $\rightarrow$       & \textit{Lint Leakage}  & 0.24                & 0.76          & 1.19 \\
    \textit{Context Bloat}, \textit{Skill Leakage}            & $\rightarrow$       & \textit{Lint Leakage}  & 0.10                & 0.75          & 1.18 \\
    \textit{Conflicting Instructions}, \textit{Lint Leakage}  & $\rightarrow$       & \textit{Context Bloat} & 0.07                & 0.67          & 1.44 \\
    \bottomrule
  \end{tabular}
\end{table*}

\section{Co-Occurrence Analysis}
\label{sec:cooccurrence}

Besides analyzing smells individually, we also investigated which of them tend to co-exist in the same \agentsmd\ files.
To discover such relationships, we used Apriori to mine association rules between the smells detected in our dataset~\cite{agrawal1996fast}.
This procedure is frequently used in software engineering research to associate code smells~\cite{Hamdi2021, Muse2020, Palomba2017}, source code files~\cite{Soto2018}, and bug types~\cite{DeSantana2024}.
In our case, we mapped each \agentsmd\ as a transaction record, and the presence of each smell in the file as an item.
For instance, a given file $F_1$ with \textit{Context Bloat} ($CB$), \textit{Skill Leakage} ($SL$), and \textit{Lint Leakage} ($LL$) is represented by the following transaction: $F_1 = \{CB, SL, LL\}$.
For the 91 \agentsmd\ files---our trans\-ac\-tions---we applied Apriori with a support of $0.05$.

Table \ref{tab:smells-apriori} presents the association rules.
The support varies between 0.06 and 0.24, i.e.,~the associated smells co-exist between 6\% and 24\% of the files in our dataset.
The confidence levels measure the probability of the consequent smell being present given the presence of the antecedent ones.
As we can see, the confidence levels are relatively high; for example, the presence of \textit{Conflicting Instructions} and \textit{Skill Leakage} increases the likelihood of \textit{Context Bloat} to 83\%.
Similar confidence appears in other rules, such as the one associating \textit{Init Fossilization} and \textit{Skill Leakage} with \textit{Lint Leakage} (83\%), and \textit{Conflicting Instructions} with \textit{Context Bloat} (81\%).

\begin{figure*}[ht]
  \centering
  \includegraphics[width=0.9\textwidth]{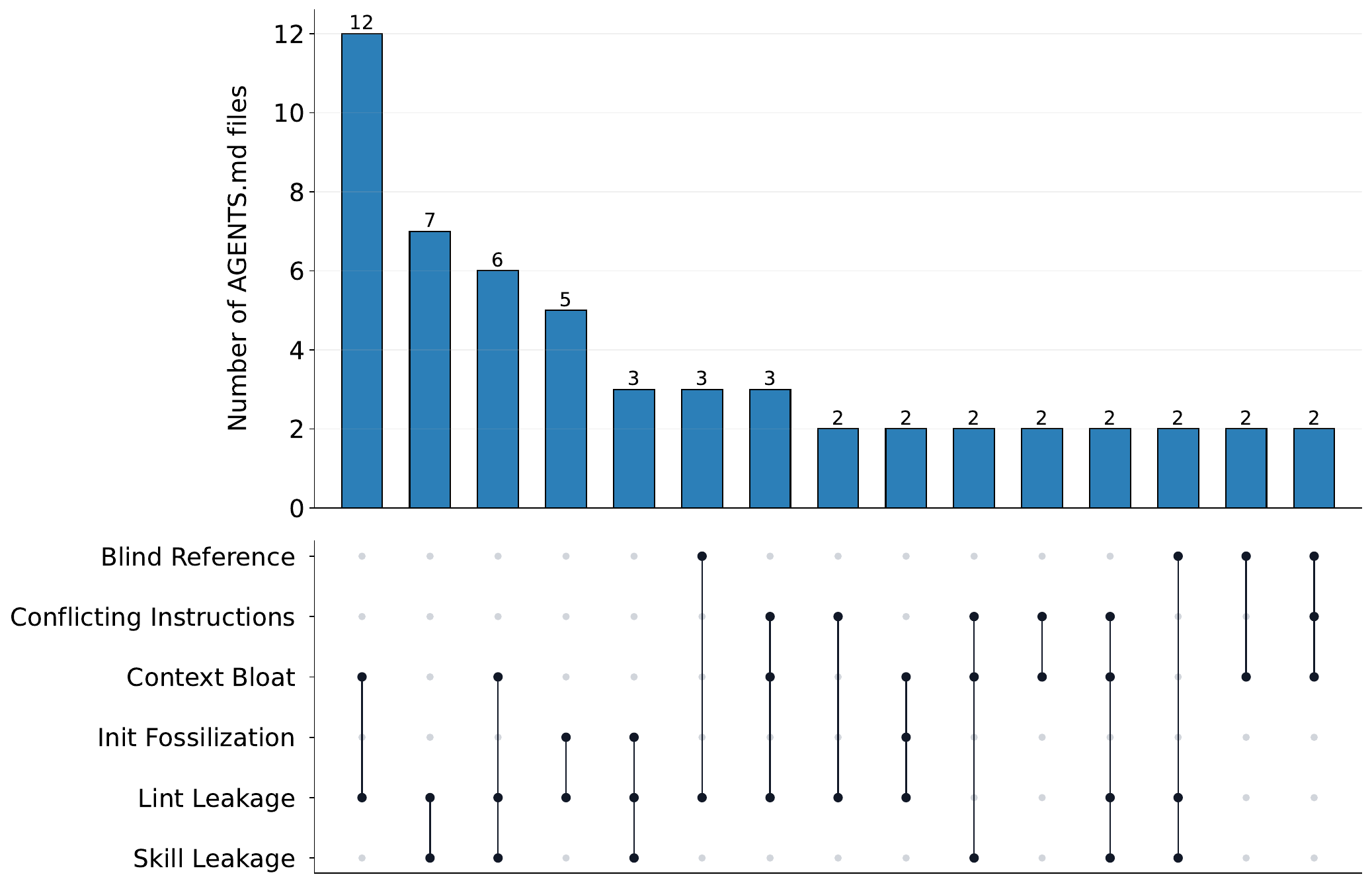}
  \caption{Smell accumulation in \agentsmd\ files}
  \label{fig:smells-upset}
\end{figure*}

Lift measures how much more likely items are to appear together than by chance.
A lift value greater than 1 indicates the smells appear in the same file more often than randomly.
We observe a strong association in two rules: \textit{Conflicting Instructions} and \textit{Skill Leakage} with \textit{Context Bloat} (1.81), and \textit{Conflicting Instructions} with \textit{Context Bloat} (1.76).
In other words, the probability of \textit{Conflicting Instructions}, \textit{Skill Leakage}, and \textit{Context Bloat} co-occurring in the same file is 1.81 times higher than if they were independent; likewise, \textit{Conflicting Instructions} and \textit{Context Bloat} are 1.76 times more likely to co-occur than by chance.

Figure~\ref{fig:smells-upset} depicts an Upset diagram showing the frequency \agentsmd\ files---bar chart on top---for each unique smell combination---intersection matrix below.
In total, we detected 15 smell combinations with at least two occurrences.
\textit{Context Bloat} and \textit{Lint Leakage} stands out with 12 occurrences; i.e.,~they appeared together in 12 configuration files.
A second group of smells shows up next: 
(a) \textit{Skill Leakage} and \textit{Lint Leakage} with seven occurrences; 
(b) \textit{Skill Leakage}, \textit{Lint Leakage}, and \textit{Context Bloat} in six files; and 
(c) \textit{Lint Leakage} and \textit{Init Fossilization} in five files.
The other 11 remaining combinations were detected in three configuration files, at most.
We observe that configuration smells frequently overlap in \agentsmd\ files, suggesting that multiple issues can compromise the performance of coding agents.

\myyellowbox{
  {\em Summary:} We leveraged six rules with co-existing smells.
  \textit{Context Bloat} is strongly associated with \textit{Conflicting Instructions} and \textit{Skill Leakage}, with lift values of 1.81 and 1.76, respectively.
  This suggests that long \agentsmd\ files often contain inconsistent or overly specific instructions.
}

\section{Threats to Validity}
\label{sec:threats}

Some steps of this work may be subject to threats to validity. In this section, we highlight and discuss these threats. \\[-0.3cm]

\noindent{\textit{Bad Practices Annotations.}}
We manually annotated the sentences from the selected documents, which can introduce some bias in the smells identified further.
To mitigate this issue, another author performed a sanity check and reviewed all annotations from the selected documents to confirm whether the content accurately represented a good or bad practice in writing configuration files. \\[-0.3cm]

\noindent{\textit{Google Search Limitations.}}
In a grey literature review, relevant results might be missed due to the specific combination of search terms. 
To mitigate this threat, we used keywords to consider both bad and good practices to expand search coverage.
% To mitigate this threat, we expanded our search queries by incorporating terms that contrast bad practices with good practices, thereby broadening the search coverage. 
Furthermore, following the guidelines from other works~\cite{garousi2018guidelinesincludinggreyliterature, lucasSmells}, we conducted preliminary search trials, adding and excluding keywords to refine our search strings. \\[-0.3cm]

\noindent{\textit{Use of LLMs for Code Smell Detection.}}
In this work, we rely on LLMs to evaluate each \agentsmd\ file and identify the presence of a specific smell. 
Large Language Models (LLMs) are highly dynamic and frequently updated by their providers. 
Another key factor is the non-deterministic nature of LLMs, which poses challenges for exact replication. 
Consequently, the performance and results of code smell detection in configuration files may evolve over time. 
To mitigate this issue, we set the model temperature to 0 and carefully analyzed the responses to ensure the identification of true positive cases.

\section{Related Work}
\label{sec:related-work}

We organize related work in three subsections: Context Engineering, Configuration Files, and Configuration Smells.

\subsection{Context Engineering}

The effectiveness of LLMs in software engineering depends on the context provided to the model. Prior work has investigated how documentation, source code, retrieved artifacts, and prompt engineering can improve model responses~\cite{nam:2024, pinto-codebuddy:2024}. Other studies have examined how developers formulate prompts and how prompting practices can be systematized for programming tasks~\cite{pister-promptset:2024, sasaki-promptpatterns:2025}. Recent benchmarks have further highlighted the repository-level nature of realistic software engineering tasks. Through SWE-bench, Jimenez et al.~\cite{swebench:2024} show that resolving real-world GitHub issues requires reasoning over repository context, modifying multiple files, and validating changes through tests. Thus, this observation motivates the study of project-level guidance for coding agents.

\subsection{Agent Configuration Files}

Recent studies have started investigating configuration files for coding agents. Chatlatanagulchai et al.~\cite{chatlatanagulchai-manifests:2025} refer to such files as \textit{Agentic Coding Manifests} and show that \claudemd\ files are predominantly action-oriented, commonly including build and run commands, implementation guidance, testing instructions, and architectural information. ~\citet{agent2026-claude-code} also analyzes \claudemd\ files from public projects and identifies recurring sections such as architecture, development guidelines, project overview, and testing.
Together, these studies show that configuration files are becoming key artifacts in agentic software development. However, they focus on describing their structure and content. In this paper, we complement this line of work by investigating misconfiguration problems in \agentsmd\ and \claudemd\ files, including excessive context, lint leakage, blind references, outdated files, and conflicting instructions.

\subsection{Configuration Smells}

Bad smells have long been studied as indicators of design, implementation, and maintenance problems, including their introduction~\cite{tufano:2015}, developers' perceptions~\cite{palomba:2014}, and manifestations across different artifacts and paradigms~\cite{depaulo:2021, taibi-microservice:2018}.
More recently, researchers have extended the smell concept to configuration and infrastructure artifacts. Rosa et al.~\cite{rosa-dockerfile:2024} investigate Dockerfile smells, i.e., violations of Dockerfile best practices that may affect reliability, security, build time, image size, and reproducibility. Urdih et al.~\cite{urdih:2026} study cache-related smells in GitLab CI/CD pipelines, proposing a catalog of ten smells and reporting that only 11\% of 228 analyzed projects were smell-free. In this paper, we also focused on domain-specific smells, but with a focus on repository-level configuration files for coding agents. Unlike Dockerfile or CI/CD cache smells, which affect build and delivery processes, smells in \texttt{AGENTS.md} and \texttt{CLAUDE.md} may directly influence how coding agents interpret project conventions, prioritize instructions, and perform development tasks.

\section{Conclusion}
\label{sec:conclusion}

In this paper, we presented a catalog of smells that may affect configuration files for coding agents, such as \agentsmd\ and \claudemd. Based on a grey literature review of 14 documents, we identified six smells and proposed heuristics to detect them in a dataset of 100 popular open-source repositories. Our results show that these smells are widespread in practice, with 91 repositories exhibiting at least one smell. In particular, {\em Lint Leakage} was the most common smell, and we also observed recurring co-occurrence patterns involving {\em Context Bloat}, {\em Skill Leakage}, and {\em Conflicting Instructions}. Since configuration files are key artifacts in agentic software development, our findings suggest that their quality deserves effort and attention. We hope that the proposed catalog will serve as a foundation for future tools and techniques to detect and prevent configuration smells in coding-agent ecosystems.

\section*{Replication Package}

The data and results of this research are available at: \url{https://doi.org/10.5281/zenodo.20600327}.

\section*{Acknowledgments}

\noindent This research was supported by FAPEMIG and CNPq.

\small

\bibliographystyle{IEEEtranN}
\bibliography{main}

\end{document}